\documentclass[fleqn,10pt]{wlscirep}
\usepackage[utf8]{inputenc}
\usepackage[T1]{fontenc}
\title{Real-time estimation of the effective reproduction number of COVID-19 from behavioral data}

\author[1,+]{Eszter Bok\'anyi}
\author[2,+]{Zsolt Vizi}
\author[3,4]{J\'ulia Koltai}
\author[2]{Gergely R\"ost}
\author[5,6,*]{M\'arton Karsai}

\affil[1]{\small Institute of Logic, Language and Computation, University of Amsterdam, Amsterdam, 1090GE, The Netherlands}
\affil[2]{\small Bolyai Institute, University of Szeged, Szeged, H-6720, Hungary}
\affil[3]{\small Computational Social Science and Research Center for Educational and Network Studies, Centre for Social Sciences, Budapest, H-1097, Hungary}
\affil[4]{\small Faculty of Social Sciences, E\"otv\"os Lor\'and University, Budapest, H-1117, Hungary}
\affil[5]{\small Department of Network and Data Science, Central European University, Vienna, A-1100, Austria}
\affil[6]{\small Alfr\'ed R\'enyi Institute of Mathematics, Budapest, H-1053, Hungary}
\affil[*]{karsaim@ceu.edu}
\affil[+]{these authors contributed equally to this work}


\begin{abstract}
Near-real time estimations of the effective reproduction number are among the most important tools to track the progression of a pandemic and to inform policy makers and the general public. However, these estimations rely on reported case numbers, commonly recorded with significant biases. The epidemic outcome is strongly influenced by the dynamics of social contacts, which are neglected in conventional surveillance systems as their real-time observation is challenging. Here, we propose a concept using online and offline behavioral data, recording age-stratified contact matrices at a daily rate. Modeling the epidemic using the reconstructed matrices we dynamically estimate the effective reproduction number during the two first waves of the COVID-19 pandemic in Hungary. Our results demonstrate how behavioral data can be used to build alternative monitoring systems complementing the established public health surveillance. They can identify and provide better signals during periods when official estimates appear unreliable due to observational biases.
\end{abstract}

\begin{document}

\flushbottom
\thispagestyle{empty}

\maketitle

\section*{Introduction}

Behavioral patterns strongly influence the outcome of an epidemic, yet observing how they change during an unfolding pandemic is among the largest challenges ~\cite{van2020using,betsch2020behavioural}. Alongside conventional survey methods, recent online and digital technologies provide new solutions to this problem. However, it is not evident how to translate large-scale observational data into actionable input for operational processes such as epidemic surveillance or modeling. Moreover, the dynamical estimation of social interaction patterns for large representative populations is problematic without entering privacy issues. We built an online/offline data collection infrastructure to continuously follow age-stratified contact matrices in a large population during the COVID-19 pandemic~\cite{karsai2020hungary}. Integrating these self-reported contact numbers from voluntarily provided anonymous online questionnaires into disease transmission models, we demonstrate how to estimate the dynamics of the effective reproduction number from behavioral data. Alongside the conventional solutions based on medical statistics and population testing, our ecosystem provides a complementary surveillance system for disease monitoring.

There are several reasons why people change the way they interact, travel, or protect themselves during a pandemic. Non-pharmaceutical interventions (NPIs)~\cite{perra2021non} such as lockdowns, school closures, mask mandates, and other regulations are the most direct causes that might induce change in people's behavior. However, fear of contamination~\cite{yildirim2021impacts}, lack of trust in governmental communication~\cite{lim2021government}, or belief in misinformation~\cite{roozenbeek2020susceptibility} can also cause a radical shift in one's social and mobility patterns, sometimes even leading to counter-effective situations like mass protests against regulations in the middle of a pandemic~\cite{kowalewski2020street}. Therefore, it is challenging to dynamically observe the convoluted effects of all these behavioral forces, not to mention their explanation by disentangled causal reasons.

It is essential to understand how people alter their social behavior~\cite{zhang2020changes,karsai2020hungary,elmer2020students} and mobility patterns~\cite{warren2020mobility,engle2020staying,leung2021real} during a pandemic~\cite{betsch2020behavioural,naughton2021health,betsch2020germany,karsai2020hungary,kittel2021austrian,manica2021impact}. These changes directly influence the way people meet, mix and interact with others, which then determines the dynamics of the disease spreading. The follow-up of direct physical contacts or proximity interactions of people are crucial from an epidemiological point of view as they provide the underlying conditions to transmit various types e.g. influenza-like illnesss~\cite{mossong2008social,wallinga2006using,zhang2020changes,singh2020age}. Therefore, the social networks of people that encode physical and proxy interactions might provide critical input to epidemic models at different levels of aggregation, like in forms of age-stratified contact matrices~\cite{prem2017projecting,zhang2020changes,mistry2021inferring,koltai2021monitoring}. Even though the dynamical monitoring of such social networks is a prime goal during epidemic crises, conventional methods like surveys and contact diaries cannot provide the necessary frequency~\cite{hoang2019systematic,fu2007contact} for their precise observations. As a solution, it is possible to exploit novel digital data collection methods with new ways of consented observations of people's social dynamics, like using online social platforms, online questionnaires, or contact tracing apps~\cite{keeling2020efficacy,koltai2021monitoring}.

One of the broadly adopted metrics to characterize the actual state of an epidemic is the \emph{basic reproduction number}~\cite{delamater2019complexity} $R_0$. This measure determines the expected number of secondary infection cases induced by a single infected individual in a fully susceptible population. If this number $R_0>1$, the number of confirmed cases will rise, whereas if $R_0<1$, there will be no sizeable outbreak. Nevertheless, during an evolving real epidemic with a large fraction of infected people, the spreading dynamics is better estimated by the \emph{effective reproduction number} $R_t$. This quantity takes the actual size of the remaining uninfected population into account and incorporates all other aspects that influence the course of the epidemic. It is affected by several factors, such as the transmission rate of the infection, the duration of infectiousness of infected individuals, or the contact frequency in the host population~\cite{dietz1993estimation}. For a given population, $R_t$ is usually calculated with statistical tools~\cite{cori_2013,wallinga2007generation} from epidemiological data like the number of fatalities or the detected number of infected cases. These numbers are collected via centralized national surveillance systems, which are not only expensive but also difficult to verify. Moreover, none of these observables provide a good solution to nowcast the actual $R_t$ values. Fatalities are usually well documented, thus, their count could potentially provide a precise measure to estimate $R_t$. However, identified COVID-19 deceased are reported usually with delays after their initial infection, due to the different course of the illness for different individuals, and also due to reporting delays. Such delays fluctuate and can mount up to weeks, which makes fatality counts impossible to use for the real-time monitoring of the epidemic. The number of detected cases are usually reported more rapidly but they provide less precise observables. These counts easily fluctuate due to extreme events or other biases. One of their most significant observational bias is caused by limited testing capacities, inducing high \emph{positivity rates}. Following the recommendation of the World Health Organization (WHO), the test positivity rate should not exceed $5\%$~\cite{world2020public} for reliable observations. However, during the early phase of the pandemic, due to the shortage of tests and later upon the emergence of highly transmissible variants, this condition was difficult to maintain. This caused severe underestimation of $R_t$ during major epidemic waves in many countries~\cite{hasell2020cross}. Other biasing factors come from case importations and local epidemic clusters, testing campaigns, or the slow data retrieval due to delayed case reporting. All these shortcomings make these conventional observables difficult to use for the precise and real-time inference of the actual $R_t$ values during an emerging pandemic. This calls for novel methods to estimate $R_t$ dynamically from alternative data sources in order to provide independent monitoring tools to follow the actual epidemic and to help operative decisions.

To answer this challenge, we have built an infrastructure that can estimate the effective reproduction number $R_t$ in real-time with remarkable precision using contact dynamics data collected online and via telephone surveys. More precisely, we collected daily age-stratified contact matrices during the first and second waves of the COVID-19 pandemic in Hungary using an online questionnaire, which was answered $538,684$ times by $235,072$ unique users since its launch. Meanwhile, we recorded the same questionnaire each month on a representative population of $1,500$ individuals via telephone surveys (for more details, see Methods). With the combination of the two datasets, we reconstructed a sequence of age contact matrices at a daily resolution, that we share through an open repository~\cite{mtxrepo} along this paper. In turn we feed these matrices as an input to a deterministic epidemic compartment model, which this way not only considers the age-stratified contact patterns of the modeled population but incorporates the effects of contact behavioral changes in its dynamics. The numerical solution of this model served us with an inferred $R_t$ function at a daily resolution, that better estimated the $R_t$ values computed from fatality rates for a reference period, as compared to case-number base estimations.

This solution provides a cheap alternative monitoring system complementing observations made via conventional surveillance infrastructures relying on the public health system. It allows for cross-validating and indicating weaknesses of official surveillance when the inference of $R_t$ is biased. Moreover, our monitoring method can closely follow the effects of NPIs on contact numbers of individuals, thus allowing us to evaluate the impact of regulations in almost real-time. In the Results section, first, we briefly describe our data collection, integration, and modeling infrastructure. Subsequently, we present our findings on the observed contact dynamics in Hungary and the reconstructed $R_t$ function that we compare to the official surveillance data. Finally, we discuss the potentials, limitations, and future directions of our results.

\section*{Results}

\subsection*{Data collection and pre-processing}

Ten days after the first officially reported COVID-19 case in Hungary, an online data collection platform was initiated to track the social and individual behavioral changes of people during the unfolding pandemic~\cite{karsai2020hungary}. The so-called Hungarian Data Provider Questionnaire ("Magyar Adatszolgáltató Kérdőív" - MASZK)~\cite{MASZKquest} data collection started on March 23, 2020 and has been continued ever since. Over this period, the online questionnaire has been answered $538,684$ times by $235,072$ unique users, which is roughly $2.4\%$ of the total population of Hungary.

\begin{figure}[ht!]
    \centering
    \includegraphics[width=\textwidth]{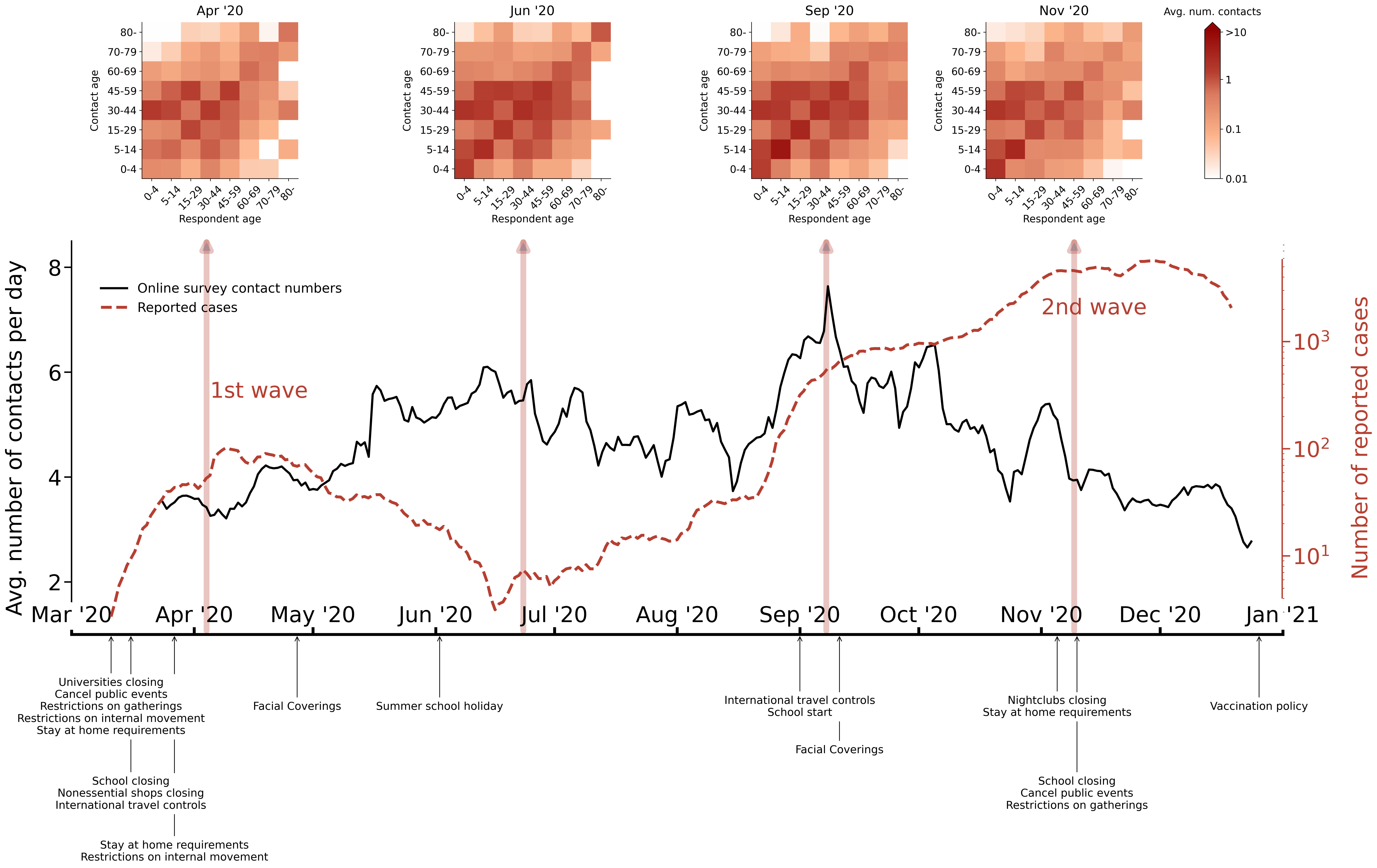}
    \caption{Average contact numbers calculated from the online survey (black line), parallel to the number of confirmed cases in the same period (red dashed line). The timeline of the most important NPI measures in Hungary is below the horizontal axis \cite{hale2021global}. Case numbers are smoothed by a 7-day sliding window, similarly to the calculated average contact numbers, that also aggregate online survey data into 7-day sliding windows. Four selected contact matrices for the 8 age groups are shown above the curves. The effects of lockdowns and school closures (or the lack of them) are evidently visible in the matrix elements.}
    \label{contactnum}
\end{figure}

Respondents were asked to estimate the number of people from eight different age groups ($0-4$, $5-14$, $15-29$, $30-44$, $45-59$, $60-69$, $70-79$, and $80+$) they got in contact with during the previous day without mask protection. Such \emph{proxy} contacts were defined as having spent more than 15 minutes within 2~m distance with someone, while at least one of them being without a mask. Relevant to this study, people also provided several of their socio-demographic characteristics (e.g. their age, gender, education level, resident municipality, etc.). Although the data collection involved only adult participants (over the age of 18), parents were asked to give their estimations about the contact numbers of their underage family members. As reference period, responses were also recorded about respondents' contact patterns from the period before the COVID-19 pandemic. In the actual study, we limit our observations to the first two epidemic waves in Hungary, falling between the 1st April and 31st December 2020, during which the same virus variant was dominantly spreading.

Although a large number of people participated in the data collection, since the online questionnaire was fully voluntary and anonymous, it did not provide a representative sample of the whole population of the country. We addressed this problem by collecting the same questionnaire in parallel using a phone-assisted survey method on a monthly basis. The interviewed population of this survey was representative for the Hungarian population along several dimensions, namely age, gender, settlement type, and education level. We summarize this data collection pipeline in the Methods section in Figure~\ref{sema} with more details on data collection, filtering, and pre-processing~\cite{koltai2021monitoring}. As a result, we could reconstruct daily contact matrices (using a statistical method explained in Methods) and follow the average number of contacts per person over the course of the pandemic, as demonstrated in Figure~\ref{contactnum} for the first two epidemic waves. While for the pre-pandemic period, we measured roughly 19.2 contacts per a person on an average day (estimated from answers between 1 April 2020 and 1 June 2020), this number drastically reduced by more than $80\%$ in 2020 March, after which contact numbers conversely followed the actual number of infected cases in the country, as shown in Figure~\ref{contactnum}.

\subsection*{Estimation of the effective reproduction number}

\begin{figure}[!ht]
    \centering
    \includegraphics[width=0.5\textwidth]{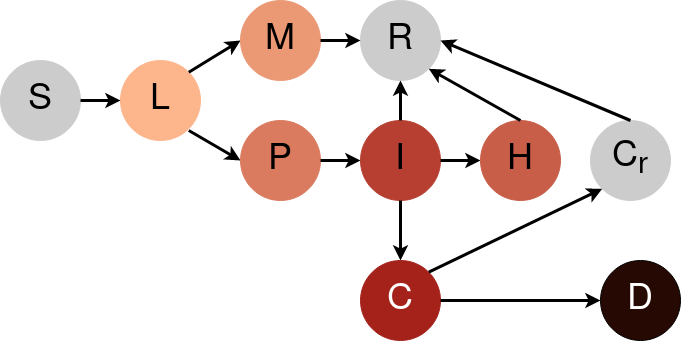}
    \caption{Compartment transmission diagram of the epidemic model}
    \label{fig:trdiag}
\end{figure}

We used the obtained daily contact matrices as an input for modeling the transmission dynamics. We estimated the time-varying reproduction number during the course of the epidemic waves by employing a deterministic compartmental model. This model contains classes for latency, infectious and hospitalized period, and relaxes the condition for homogeneous mixing via tracking transmission routes between age groups in the population. For the visual representation of all transitions between the compartments, see Figure \ref{fig:trdiag} and for the system of resulted equations, see \ref{eq:seir}. To incorporate age-stratified transmission patterns, we used the previously computed dynamical daily contact matrices. They represent the heterogeneity of the social contacts among individuals of different age groups, thus, they form the basis for the calculation of the associated effective reproduction number. Further, we considered seasonality effects deeming periodically lower transmission rates of the epidemic during the summer periods. The model includes an age-dependent parameter for susceptibility, which is smaller for young individuals implying less effective transmission. For details about the compartment structure, parametrization, and seasonality integration of the epidemic model, see Supplementary Information.

We iteratively solved the system defined by the model, starting from the state of the previous day and replacing the contact matrix of the next day in the simulation. During the model solution, the time-dependent age vectors of susceptible individuals were used to calculate the effective reproduction number on a daily basis. Since the number of infections during the first wave in Hungary was very low, the significance of tracking the depletion of susceptibles in age groups appeared only in the second wave. Consequently, we started our modeled epidemic from a fully susceptible population in April 2020 and simulated it for nine months, until January 2021, which corresponded roughly to the end of the second pandemic wave in Hungary. We chose a reference period, when the case number based $R_t$ estimate is deemed accurate, and used this reference point to calibrate the relation between social contacts, fraction of susceptibles, and $R_t$ value. The selected reference period was mid-September, since for this period we obtained the most reliable $R_t$ estimate, which was confirmed by different types of epidemiological data, such as incidence data, hospitalizations and mortality trends (see~\cite{oroszi2021az}). Having this reference point fixed, following our methodology, we could calculate the $R_t$ rates for periods prior and posterior to the reference point. As a model output, we computed the $R_t$ effective reproduction number using the so-called Next Generation Matrix (NGM)~\cite{DiekmannHesterbeek,RostViruses} method, which partitions the model structure to transition (focusing on the flow between the classes) and transmission (involving age-specific social patterns) parts. At a time point $t$, we compute a matrix whose dominant eigenvalue provides the value of $R_t$. For complete description of the methodology, see the Supplementary Information.

Note, that in a GitHub repository we share the epidemic simulation code incorporating the dynamical contact matrices~\cite{mtxrepo}.

\begin{figure}[ht!]
    \centering
    \includegraphics[width=\textwidth]{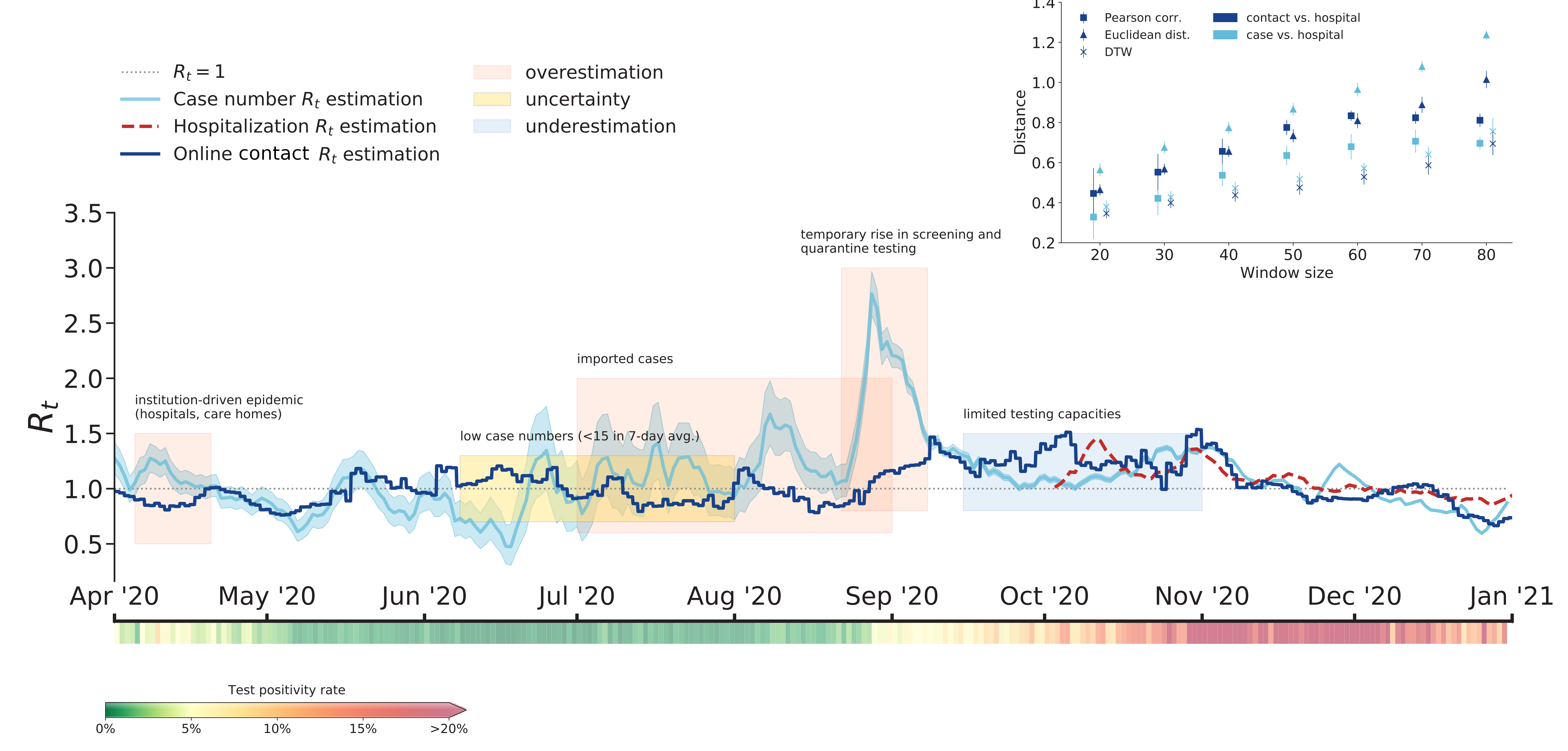}
    \caption{Effective reproduction numbers between 1st April 2020 and 31st December 2020 in Hungary estimated from the daily contact matrices of the online survey (dark blue), and from the case numbers using the Cori method (light blue) \cite{cori_2013} with statistical confidence intervals shown as blue shaded area. Reference $R_t$ estimated from the 3rd October 2020 using hospitalization numbers is shown by a red dashed line. The black dotted line indicates the $R_t=1$ critical reproduction number. Colored stripe below the horizontal axis depicts the test positivity rate as a percentage of positive tests of all tests taken in the country on the given day. Annotated boxes show periods where methods based on case numbers either overestimate (red) or underestimate (blue) the reproduction number, and where the method exhibits uncertainty due to very low case numbers (yellow). The inset presents the comparison of the contact matrix based and case number based $R_t$ estimations to the reference curves based on hospitalization numbers to estimate $R_t$. Differences between curves were measured by the Pearson correlation as a similarity, and Dynamic Time Warping and Euclidean distance as distance metrics with $95\%$ confidence intervals shown.}
    \label{rt}
\end{figure}

\subsection*{Alternative reproduction number surveillance for Hungary}

The estimated effective reproduction numbers are shown in Figure~\ref{rt} during the first two pandemics waves in Hungary. There, the dark blue curve corresponds to the $R_t$ estimated from our model solutions, which relies on online data and it takes into account the dynamical change of contact patterns. On the other hand, during the same period, several other methods have been proposed and applied to track the effective reproduction number in real time~\cite{wallinga2007generation,cori_2013}. These estimations commonly rely on the reported case numbers, which suffer from numerous biases, which could even change during the pandemic. We use one such estimate publicly available at~\cite{ferenci2022rt} that we indicate by a light blue curve and the corresponding $95\%$ confidence interval in Figure~\ref{rt}. This curve represents an estimation of $R_t$ computed by the Cori method~\cite{cori_2013} using the official Hungarian case numbers.

By looking at Figure~\ref{rt}, both the official and simulated $R_t$ values were smaller than one in the spring of 2020, confirming that the first wave was successfully suppressed. It was hovering around one during the summer and started grow distinctly above one from September onward, signifying a large second wave. The $R_t$ value dropped below one at the end of November, marking the peak of the second wave, and remained below one afterwards, indicating the decay phase of the second wave. Generally, the effective reproduction numbers estimated from online data and model simulations were following surprisingly well the officially reported $R_t$ numbers for the entire period. Moreover, given a past reference point, this method allows us to make $R_t$ estimates not only retrospectively, but also in real time, presuming that social mixing data is collected continuously in real time as well.

At the same time, it is evident that the two estimated $R_t$ curves deviate from each other during some periods. We indicate these periods with colored boxes, during which the official reproduction number was deemed less reliable and deviated from the modeled curves. Following a chronological order, the first wave in Hungary in the spring of 2020 was dominated by outbreaks in healthcare and social care institutions. Therefore, these outbreaks generated a sharp increase in reported cases, leading to some short living spuriously high $R_t$ values in the case number based estimate. Yet, although these cases increased the number of confirmed cases, these high values did not represent the spread of the infection in the general population, as they correspond to well-contained local outbreaks~\cite{RostViruses}. This was captured by the modeled $R_t$ values, which remained under one during this period.

Subsequently, in mid summer 2020, reported values have been noisy due to low case numbers (<10, yellow period in Fig.~\ref{rt}). This explains the very low numbers of case based $R_t$ numbers, as compared to the modeled values, which remained higher due to the relatively large number of social contacts during the summer. After lifting the border closure measures, during the late summer, there was a period of time when the infection numbers were driven by case importations from abroad, inflating again the $R_t$ estimate above the modeled values. From mid august 2020, the government carried out a large screening campaign in freshmen camps before the start of the higher education autumn semester. This has lead to an artificial peak in the infected case number based curve. Meanwhile, the convoluted effects of case importations, mass events like weddings and freshmen camps, the increased social contact numbers due to the beginning of the school year, and the seasonally augmented transmission rates led to the emergence of the second wave in Hungary. This was actually well reflected by the modeled curve using contact numbers that signaled increasing $R_t$ numbers from 2020 September.

In the exponential phase of the second wave, Hungary quickly reached its limit in testing capacity, and the reported case numbers did not grow any further. This resulted in a misleadingly low $R_t$ estimate in the case number based curve~\cite{oroszi2021az} in October 2020. This phenomenon is especially striking in the period when the test positivity rates, indicated by the colored stripe below the horizontal axis of Figure~\ref{rt}, grew steadily from the beginning of September until November (blue period on the main panel). Based on the estimation of $R_t$ derived from case numbers, public health authorities did not assess the pandemic situation correctly in this period, which delayed the introduction of more serious NPI measures to control the fast spreading. Interestingly, from our alternative surveillance, we observed more realistic $R_t$ numbers that were significantly higher than one during this period. The case number based and online estimated $R_t$ curves matched again once the test positivity rate reached a stationary value around the stagnation period of the second pandemic wave from November 2020. Consequently, if we compare our contact number based $R_t$ estimations to the case number based approach, we can see that in the indicated periods that suffer from one of the aforementioned problems, we have a better estimate than the reference curve. 

\subsection*{Validation of inferred reproduction numbers}

Apart from the previously cited limitations, the case number based $R_t$ estimations also suffer from the time lag that comes from the disease course (most cases turn positive once the patients have symptoms), and the delays in sample processing and test reporting procedures. The identification and correction of the biases in these case number based estimations require tremendous epidemiological work, high quality data on each individual case beyond raw case numbers, and an intimate knowledge of the country's surveillance and reporting system. On the other hand, hospitalization numbers, Intensive Care Unit (ICU) admission rates, or the number of deaths have even larger time lags due to the temporal disease progression. Nevertheless, these numbers are more reliable and indicative of the spreading than the reported number of confirmed cases. 

We use such an $R_t$ curve, estimated from daily hospitalization counts, to validate whether the contact number based or the case number based $R_t$ curves meet closer with the reference. The hospitalization number based $R_t$ curve (red dashed line in Fig.~\ref{rt}) was collected only after 2020 October, as data from earlier periods are not available. We performed pairwise comparisons between the case number vs. hospital number based and the online contact numbers based vs hospital number based curves. To compare these temporal sequences we used multiple metrics: the Pearson Correlation as a similarity measure, and the Euclidean Distance and the Dynamical Time Warping as distance measures. Comparisons were made by using sliding time windows with different sizes, indicated as the x-axis scale in Fig.~\ref{rt} inset. There we see that the contact number based $R_t$ curve is significantly more similar for any window size to the hospitalization based reference curve as compared to the similarity of the case number based estimates.

Although we could demonstrate that the contact number based $R_t$ estimates approximate the reference values better, our goal with these proposed methodology was not to replace official surveillance results using case numbers for their estimates. We rather aimed to propose alternative surveillance observations that complement the official monitoring tools. Remarkably, in the 2020 autumn period of growing test positivity rates, our estimation remains above one, indicating a fast growing epidemic. This highlights an important aspect of our methodology provided by monitoring social mixing dynamics, as it allows to overcome some of the biases in the case number based $R_t$ estimations. Interestingly, we can provide an $R_t$ value estimate, which during biased intervals give a better picture about the unfolding epidemics, this way complementing the traditional surveillance system.

\section*{Discussion}

Beyond official surveillance relying on detected case numbers and medical statistics, alternative methods can monitor the unfolding of a pandemic~\cite{kostkova2021data}. Some methodologies rely on geo-localized web search and social media tracking to nowcast trends in epidemic-related topics~\cite{dugas2013influenza,tang2018social}. Despite the popularity of these methods, their vulnerabilities and limitations got evident over the years~\cite{lazer2014parable}. In several other studies, the reproduction number of an epidemic is estimated from the mobility patterns of people. Human mobility followed by mobile phone activities, GPS devices, or check-in data could signal the traveling, commuting, and mixing patterns of people, which largely determine the spread of an epidemic in a larger population. However, despite the many promising results~\cite{vanni2021use,jung2021predicting,gozzi2022anatomy}, the mobility activity of people, quantified by various indices~\cite{bao2020does, szocska2021countrywide,wang2020examining}, does not always follow the epidemic curve of the pandemic. People accept and follow some interventions better, while some others less. Whereas mask use became a worldwide accepted norm, mobility restrictions became less and less enforced and followed. Therefore, the trends of people's mobility and the number of infections may diverge~\cite{gottumukkala2021exploring, bokanyi2021kontaktkutatas}. Also, statistics, such as the age-stratified mixing patterns or the fraction of recovered population, are hard to follow with mobility data, which prevents the precise estimation of the effective reproduction number using this type of data sources. For all these reasons, although mobility monitoring plays an essential role in estimating mixing patterns, it may appear as a less correlated direct indicator of epidemic prevalence over time. Our modeling approach could provide a more reliable solution, as it integrates dynamical contact information into epidemic models in the form of time-varying age-stratified contact matrices. This way, it directly introduces the effects of interventions and behavioral changes through the recorded dynamics of social interactions, which leads to better approximations of possible transmission events of disease spreading.

Nevertheless, our proposed methodology has certain limitations. Most importantly, it heavily relies on the respondent population size and its representative composition (for the change of representativeness of our data see Supplementary Information). Although by using combined online/offline data collection methods, we accounted for the non-representativeness of the recorded data, this remains a challenge. We found the representative weighting dimensions robust over the observation period, but they may change over time. Thus, repeated data collection campaigns via representative telephone surveys are necessary. At the same time, while we account for seasonality, other environmental factors (like humidity or pollution level) may influence the epidemic outcome. On the other hand, due to the continuous evolution of new genetic variants, although the biological profile of the pathogen (e.g. its transmission rate or the length of the incubation period) may change, that can be considered in our model. We could also incorporate the dynamics of vaccination and waning immunity into our modeling framework. Finally, voluntary responses may suffer from cognitive distortion, potentially inducing over-representative numbers of answers from overly alert people during some time of the pandemic. Such biases are difficult to capture with our demographic variables in the representativity correction process.

The dynamically varying number of social contacts are one of the primary indicators of social mixing that can potentially estimate the transmission rate of an influenza-like illness. To demonstrate this approach, we described a data collection effort to record age-stratified contact matrices in Hungary at a daily resolution. We integrated them into a deterministic compartment model to estimate the temporal evolution of the reproduction number of the COVID-19 epidemic. This innovative solution provides a cheap and near real-time surveillance system independent of public health data. Instead of using contact tracing, frequent representative surveys, or medical statistics, it relies on the combination of alternative data sources collected online and offline with the involvement of thousands of individuals. It provides a powerful solution to cross-validate results from conventional surveillance systems or to identify biases or uncertain estimation periods.

The overall goal of NPIs is to suppress the possible epidemic transmission by decreasing the number of contacts of people through different ways of regulations. Our framework provides a way that can directly monitor the effectiveness of these restrictive measures. It allows to immediately evaluate their impact on larger populations compared to behavioral patterns before and after the regulated period. This method provides an inventive tool for disease monitoring with easy implementation in many countries. Beyond its scientific merit, it may provide effective monitoring of the consequences of national interventions, to follow the effects of population-level behavioral changes, and to inform intervention planning and policy design. Moreover, as we demonstrated in the case of Hungary, it allows to complement traditional surveillance systems in two ways: by signaling periods when official monitoring infrastructures are unreliable due to observational biases; and by providing more accurate signals of the epidemic dynamics during these periods. For all these reasons this methodology should be integrated into future public health surveillance systems for more precise epidemic monitoring.

\section*{Methods}

\subsection*{Data collection and reconstruction pipeline}

\subsubsection*{Online data collections}

We collected data via an online questionnaire~\cite{maszk:team} that users could fill using web browsers or mobile phone apps. Our data collection was completely anonymous using local encrypted browser cookies to improve user experience, without requiring participants to share any personal identifier that could be used for their identification. The data collection was fully complying with the actual European and Hungarian privacy data regulations and was approved by the Hungarian National Authority for Data Protection and Freedom of Information~\cite{naih}, and also by the Health Science Council Scientific and Research Ethics Committee (resolution number IV/3073- 1 /2021/EKU). During our analysis all methods were performed in accordance with these relevant guidelines and regulations.

The responses contained information on the demographics and family structure of the anonymous users, their contact numbers from the previous day by the age of the contacted people in different situations (e.g. indoors, outdoors, at the workplace etc.), and other questions relevant to their behavior during the epidemic (see~\cite{koltai2021monitoring} for further description of the questionnaire). To define what counts as a contact, as explained in the questionnaire, we considered two people to be connected if they spent at least 15 minutes without mask protection at a distance less than 2~m (proxy contacts) from each other. Household members were automatically counted as contacts (family contacts) using the family members' age to consider them in the age-contact matrix. Although data was not directly collected about children, adults (typically parents) could fill a special part of the questionnaire to give their estimations about the proxy contact numbers of their underage family members (typically children).

\begin{figure}[h!]
    \centering
    \includegraphics[width=\textwidth]{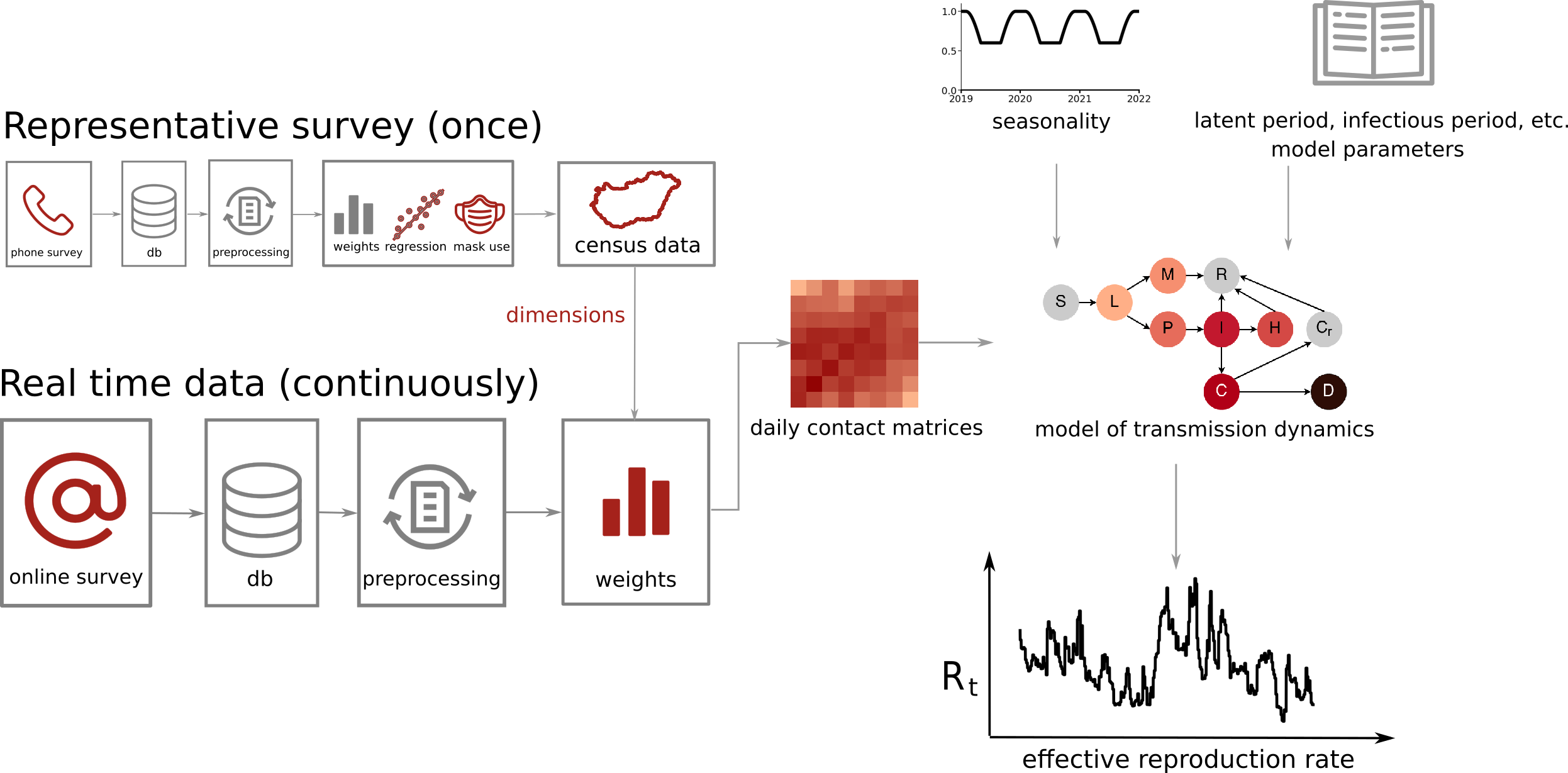}
    \caption{Schematic diagram of the data collection, data processing, and modeling pipelines. The representative phone survey is used for calculating the most important demographic dimensions that influence the average contact numbers of people, and for estimating mask use percentage in different age groups of children. Daily contact matrices are created using a 7-day sliding window from the online survey, adding user weights to correct for sample representativity using the relevant demographic dimensions and their population distribution from official census statistics ~\cite{census:2011, microcensus:2016}. Daily contact matrices are then used as input parameters to the compartmental model that uses also biological and medical parameters, as well as a seasonality correction function for the estimation of the daily effective reproduction number $R_t$.}
    \label{sema}
\end{figure}

Our observations were focusing on the period from 26/03/2020 to 31/12/2020, during which we collected $429,267$ responses from $230,878$ unique users. To avoid high noise rates, we aggregated daily online answers with a 7-day sliding window that we shifted by one day through the observation period. To make the answering more comfortable for the respondents (and thus, to increase the willingness for participation), we provided intervals for their estimated number of contacts, that we converted to their midpoints before calculations (category conversions were 0:0, 1-2:1, 3-6:4, 7-15:11, 16-30:23,31-60:45, 60+:80). We used these numbers to estimate 8$\times$8 age contact matrices from the online data on a daily basis. For a detailed schematic representation of the data collection pipeline see Figure~\ref{sema}.

\subsubsection*{Representative data collection}

As the participation in the online data collection was voluntary, respondents were not representative for the whole population of the country, moreover, their composition could change on a daily basis (see SI Figure~\ref{fig:population}). To account for these shortcomings and to record a representative sample, we started a smaller scale data collection campaign with different methodology but using the exact same questionnaire. This survey has been conducted with CATI (Computer Assisted Telephone Interview) survey technique by a public research company. The data collection started in April 2020 and has been repeated monthly. The respondents were selected by a multi-step stratified probability sampling technique from a database containing both mobile- and landline phone numbers. The sample is representative for the Hungarian adult (18 years old or older) population in terms of gender, age, education level and type of settlement; sampling errors were further corrected by post-stratification weights. Depending on the month, the numbers of recorded complete responses in each wave of the data collection were between $1,000$ and $1,500$, which fits the standard size of representative surveys in Hungary. The overall response rate was relatively high, $\sim 49\%$ as compared to other similar size surveys. In comparison, according to the data collection company, the average response rate of similar data collection methodologies at a nationally representative survey is between 15-20 percent. The collection of one wave generally took one week, where two-third of the responses corresponded to weekdays, and one-third to weekend days. Although telephone survey data has been collected once per month only with smaller sample size as compared to the online survey, it provided us with generalizable information about the contact patterns of the Hungarian adult population. 

Taking the collected raw data we built up a data-cleaning pipeline to prepare the data for further analysis. This pipeline has been applied on both online and representative data. First, to avoid skewed averages due to outliers, we filtered survey answers if they contained very high out-of-home total proxy contact numbers added up for all age group. In this case we chose to drop the top 0.5 percentile of total contact numbers corresponding to a cut at more than 90 proxy contacts. Moreover, in the online survey we also omitted the answers from the analysis if the contact numbers have been larger than the average plus two standard deviations within the respondents' own age group within the given time window or in the representative survey. In the exceptional cases when the number of responses within a time window for one age group was insufficient to calculate the standard deviation, we took an age-independent upper threshold computed from the system average. The latter filtering process was necessary for the online age-stratified matrices, since the $R_t$ calculation that was based on the spectral radius of the Next Generation Matrix method was very sensitive to sparse elements, and global filtering was unable to capture age group dependent outliers.

Because online responses for children did not contain information on their mask use, we estimated their contact numbers by re-scaling their reported contact numbers using their mask use percentages based on age and contact numbers from the representative survey of October 2020. Contact number is an important factor in this variable, because children tended to use masks in higher percentages in more crowded settings such as classrooms. Table~\ref{tab:kid_mask_use} shows the mask use correction factors from the representative survey applied to the online responses of children. 

\begin{table}[]
    \centering
    \begin{tabular}{lrr}
        \textbf{Age} &  \textbf{0-22 contacts} & \textbf{23+ contacts}\\
        3-6 & 0.1324 & 0.1509\\
        7-10  & 0.2720 & 0.5635\\
        11-14 & 0.3665 & 0.6979\\
        15-17 & 0.3665 & 0.7100
    \end{tabular}
    \caption{Mask use fractions of children based on age and contact number from representative survey}
    \label{tab:kid_mask_use}
\end{table}

\subsubsection*{Contact matrix reconstruction}

To account for the non-representative biases in the online data, we worked out a method to dynamically estimate weights for each online respondent to re-weight the online data to create a close-to-representative population. For the selection of the weighting dimensions our goal was to identify socio-demographic variables with available population-level distributions, which were also present in our online questionnaire. First, we tested which variables affect the proxy contact numbers of the respondents using the representative survey. As the proxy contact numbers of the respondents can only be non-negative integers, we used negative binomial regression models on the first two waves (conducted in April and May, 2020) of the representative data collection together. Based on this model (see results in SI Table~\ref{table:negbinom}), we identified age, highest education level, region, type of settlement, and the interaction of gender and work status as significantly affecting the number of proxy contacts. For the validation of the weighting dimensions see Supplementary Information.

Using these variables and the latest census data~\cite{census:2011, microcensus:2016}, we calculated $w^x$ weights for every user $x$ using the iterative proportional fitting method~\cite{bishop2007discrete} in each 7-day time window. This weighing methodology adjusts the cells of a contingency table created by the empirical distribution of the weighting dimensions in a way that their marginals fit to the expected distribution of the same dimensions. Empirical distributions were taken from the online survey, expected distributions were provided by the census data. One of the main advantages of this weighting methodology compared to standard cell weighting is to induce less likely extremely high or low weights - which could make the estimations unstable~\cite{lavrakas2008encyclopedia}. Thus, the actual weighted user sample within a one-week daily sliding window had marginals fitted to official census marginal distributions along the selected variables. We summarize this data construction pipeline in Figure~\ref{sema}, while for a detailed description of a similar regression choice we refer to \cite{koltai2021monitoring}. 

Finally, to construct age-stratified contact matrices for each period, we categorized each respondent into eight age groups, namely $0-4$, $5-14$, $15-29$, $30-44$, $45-59$, $60-69$, $70-79$, and $80+$. We constructed $8\times 8$ matrices with column indices corresponding to the age group of the respondents and row indices correspond to the age group of their contacts. To formally define this matrix on the population level we follow the same procedure as described in~\cite{koltai2021monitoring}: Let assign by $X$ be the set of respondents (ego), and by $Y$  the set of individuals who are contacts of some $x\in X$. For a specific $x$, let $N_x \subset Y$ be the set of individuals who are contacts of $x$. We assign by $a(x)\in A=\{1,\dots,8\}$ the age group of an individual $x$. We define the matrix $M^{x,y}$ for each $x\in X$ and $y \in N_x$ as $\left(M^{x,y}\right)_{i,j}=1$ if $a(x)=j$ and $a(y)=i$, and zero otherwise. For an ego $x$ we can now compute its individual contact matrix as $\mathrm{M}^x=\sum_{y\in N_x} M^{x,y}$. Finally, we use an individual weight $w^x$ assigned to each ego, coming from the IPF weighting method described above. This weight effectively describes how much an ego and its contacts should be considered in order to receive a contact matrix for a closer-to-representative population. Finally, the population level contact matrix is computed by $\textbf{M}=\sum_{x \in X} w^x \mathrm{M}^x \big/ \sum_{x \in X} w^x$.

\section*{Data availability}

In this repository~\cite{mtxrepo} we share all code and data necessary for the reproduction of our results. The shared data incorporates the source code for epidemic simulations and the data recording the empirical dynamical contact matrices. Other datasets are openly available as referenced in the text.

\bibliography{refs}

\section*{Acknowledgements}
This work was completed in the National Laboratory for Health Security of Hungary. The authors are grateful for T. Ferenci to compute and share the reference data and for F. Bartha for his contributions in epidemic modelling. The authors are thankful for A. Vespignani for the insightful comments and for N. Samay for visualization advice. We acknowledge support from the framework of the Hungarian National Development, Research, and Innovation (NKFIH) Fund 2020-2.1.1-ED-2020-00003. J.K. was supported by the Premium Postdoctoral Grant of the Hungarian Academy of Sciences. M.K. is thankful for the support from the DataRedux (ANR-19-CE46-0008) project funded by ANR, the SoBigData++ (H2020-871042) project and the EmoMap CIVICA project. G.R. was supported by NKFIH FK 124016. ZS.V. and G.R. were supported by the Ministry of Innovation and
Technology of Hungary from the National Research, Development
and Innovation Fund, project no. TKP2021-NVA-09.

\section*{Author contributions statement}
J.K., M.K. and G.R. conceived the survey data collection, E.B., J.K. and M.K. analyzed the data, ZS.V. E.B. and G.R. carried out the model computations. All authors designed the research, wrote and reviewed the manuscript.

\section*{Additional information}

\subsubsection*{Competing interests}
The authors declare no competing interests.

\subsubsection*{Materials \& Correspondence}
Correspondence and material requests should be addressed to M. Karsai (\url{karsaim@ceu.edu}) and G. R\"ost (\url{rost@math.u-szeged.hu}).

\cleardoublepage

\section*{\Large Supplementary Information}

\section*{Validation of the weighting dimensions}

To identify the main weighting dimensions for the iterative proportional fitting, Table~\ref{table:negbinom} shows the results of the negative binomial regression model on the proxy contact numbers of the respondents. Here the independent variables are the dimensions with available population-level distribution, and which were also present in the questionnaire. We can observe that in each dimension, there is at least one category that significantly affects the proxy contact numbers compared to the reference category. These results suggest that all dimensions from this regression model should be considered as weighting dimensions on the online data since they all influence the contact patterns of people.

\begin{table}[h!]
    \centering
    \begin{tabular}{l|r|r|r|r}
        \textbf{Independent Variables} &  \textbf{B} & \textbf{Std. Error} & \textbf{Wald Chi-Square} & \textbf{Significance}\\
        Intercept & 0.270 & 0.098 & 7.61 & 0.006\\
        &   &   &   &  \\
        19-29 years old & 0.846 & 0.072 & 137.24 & 0.000\\
        30-44 years old & 0.983 & 0.063 & 242.37 & 0.000\\
        45-59 years old & 0.964 & 0.062 & 240.67 & 0.000\\
        ref: 60 years old or older &   &   &   &  \\
        &   &   &   &   \\
        max. vocation & 0.078 & 0.062 & 1.58 & 0.209\\
        secondary education & 0.212 & 0.058 & 13.17 & 0.000\\
        ref: higher education &   &   &   &  \\
        &   &   &   &  \\
        other region & 0.143 & 0.066 & 4.72 & 0.030\\
        ref: Central Hungary &   &   &   &  \\
        &   &   &   &  \\
        Budapest & -0.096 & 0.083 & 1.34 & 0.248\\
        county town & -0.167 & 0.064 & 6.74 & 0.009\\
        city & -0.128 & 0.052 & 6.16 & 0.013\\
        ref: village &   &   &   &  \\
        &   &   &   &  \\
        man works & 0.756 & 0.055 & 189.55 & 0.000\\
        man does not work & 0.386 & 0.066 & 33.99 & 0.000\\
        woman works & 0.661 & 0.063 & 111.57 & 0.000\\
        ref: woman does not work &   &   &   &  
        \end{tabular}
\caption{
    Negative binomial regression model on the number of contacts with the potential weighting dimensions as independent variables
}
\label{table:negbinom}
\end{table}

\section*{Change in the population composition}

\begin{figure}[ht!]
    \centering
    \includegraphics[width=0.8\textwidth]{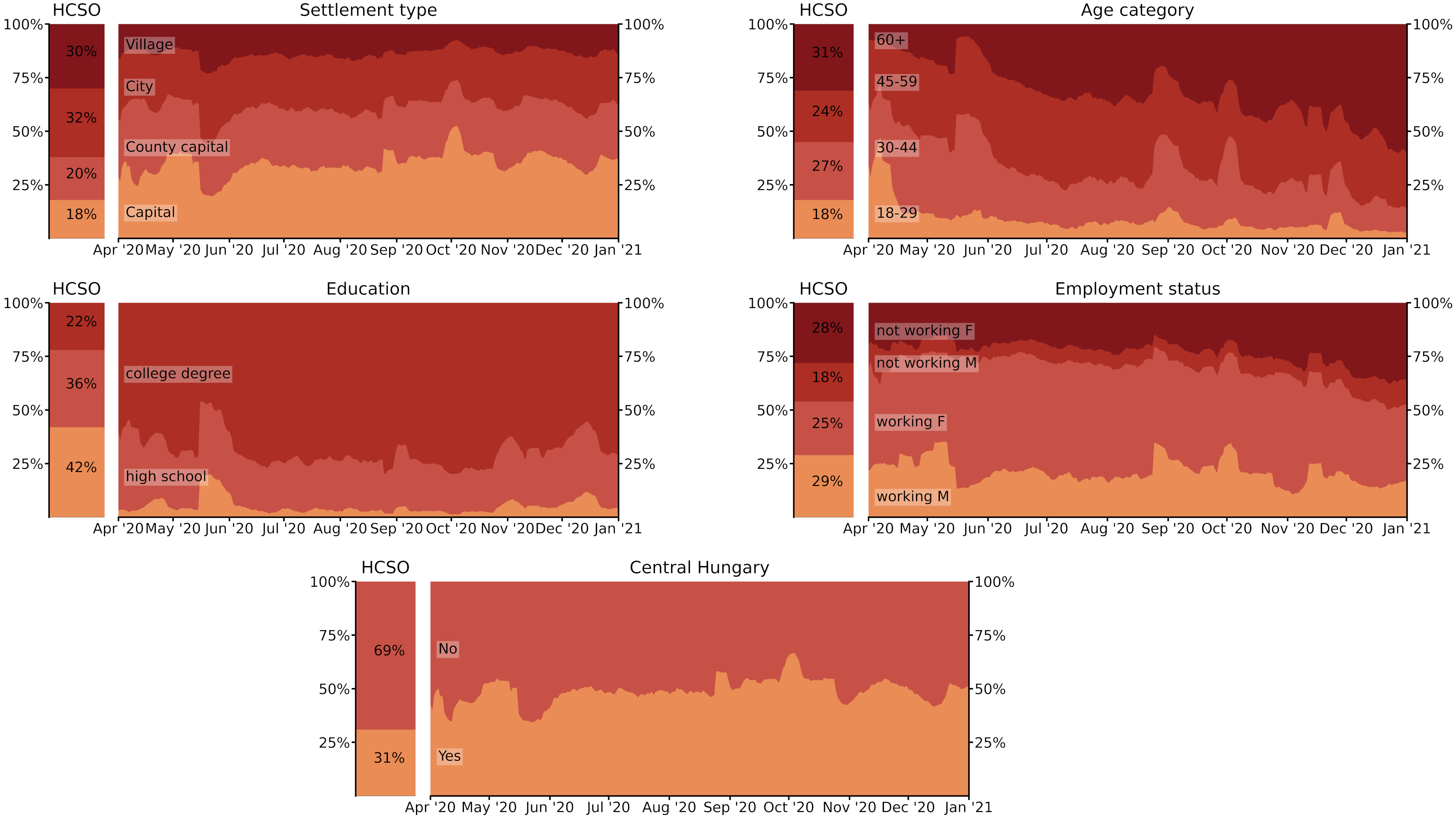}
    \caption{Distributions of users with respect to the demographic attributes used for the IPF weighting in the online survey. Reference percentages representative of the adult population obtained from official census data~\cite{census:2011, microcensus:2016} are to the left of each subplot labeled as HCSO (Hungarian Central Statistical Office). Central Hungary refers to whether a user lives in the EU region of the capital city, Budapest, and its surrounding county, called ``Pest megye''. Employment status is measured for both genders, M=male and F=female.}
    \label{fig:population}
\end{figure}

In the online survey, users were not representative of the population of Hungary. Moreover, their composition could change on a daily basis. Figure~\ref{fig:population} shows the composition as a function of time throughout the time period of this analysis alongside with the representative percentages obtained from official statistics of the Central Bureau of Statistics of Hungary (KSH)~\cite{census:2011,microcensus:2016}. Variables correspond to the significant dimensions that have been determined from the representative survey regressions, and that have the largest influence on people's contact behavior.

\section*{Epidemic model}

For investigating the dynamics of the COVID-19 epidemics, we use a slightly modified deterministic model from \cite{RostViruses}. In the following, we briefly introduce the mechanisms of the epidemic model. This model is defined on a population of people where we denote by $S$ the susceptibles, i.e. who can contract the disease. If individuals who get contracted the disease first get latent ($L$), i.e. carry the virus, but they have no symptoms yet. Then a large fraction of the latents transit to the class for asymptomatic cases, i.e. having at most mild symptoms (denoted by $M¢$), but with the ability to infecting susceptible individuals. Others develop more severe symptoms, they proceed first to the pre-symptomatic ($P$), then to the infected compartment ($I$). Individuals from $A$ compartment will all recover and consequently proceed to class for recovered (denoted by $R$), while symptomatically infected individuals may either recover without requiring further treatment or become hospitalized.

As we have seen in the COVID-19 pandemic, it is of high importance to be able to estimate number of hospital beds and intensive care unit (ICU) beds, thus we differentiate symptomatically infected individuals who need hospital and critical care (ICU), denoted by $H$ and $C$, respectively. We assume that patients admitted to non-intensive treatment will all recover, thus proceed to class $R$, however, fatal outcome may occur for individuals from class $I_c$ implying the transition from $C$ to the $D$ compartment. Those who are out of ICU and on the path to recovery are first collected in the compartment $C_r$ from where they proceed to class $R$.

Additionally, we assume that the latency and infectious periods are gamma distributed, and for their modeling we use so called the linear chain trick, i.e. we divide classes $L$, $A$ and $I$ into further compartments. For gamma distributed latency period with Erlang parameter $m=2$, we introduce classes $L_1$ and $L_2$, for infectious period with Erlang parameter $m=3$, for asymptomatically and symptomatically infected individuals we have $M_1, M_2, M_3$ and $I_1, I_2, I_3$ compartments, respectively.

Considering all the introduced compartments, the dynamics of the system is described by the following differential equation system:

\begin{equation}
\label{eq:seir}
\begin{split}
S^{a\,\prime}(t) = {} & -    
    \frac{S^{a}(t)}{N^{a}(t)} \sigma^a
        \sum_{k \in \{1, \ldots, 8\}} 
        \left[ 
            \beta_\mathbf{p}^{(k, i)} I_\mathbf{p}^{k}(t) + 
            \sum_{
                \mathbf{j} \in 
                \{\mathbf{m}, \mathbf{i}\} \times \{1, 2, 3\}} 
            \beta_\mathbf{j}^{(k, i)} I_\mathbf{j}^{k}(t) 
        \right], \\
L_\mathbf{1}^{a\,\prime}(t) = {} & 
     \frac{S^{a}(t)}{N^{a}(t)} \sigma^a
        \sum_{k \in \{1, \ldots, 8\}} 
        \left[ 
            \beta_\mathbf{p}^{(k, i)} I_\mathbf{p}^{k}(t) + 
            \sum_{
                \mathbf{j} \in 
                \{\mathbf{m}, \mathbf{i}\} \times \{1, 2, 3\}} 
            \beta_\mathbf{j}^{(k, i)} I_\mathbf{j}^{k}(t) 
        \right]
        - \alpha_{\mathbf{l}, \mathbf{1}}^{a} L_\mathbf{1}^{a}(t), \\
L_\mathbf{2}^{a\,\prime}(t) = {} & 
    \alpha_{\mathbf{l}, \mathbf{1}}^{a} L_\mathbf{1}^{a}(t) - 
    \alpha_{\mathbf{l}, \mathbf{2}}^{a} L_\mathbf{2}^{a}(t), \\
I_\mathbf{m, 1}^{a\,\prime}(t) = {} & 
    p^{a} \alpha_{\mathbf{l}, \mathbf{2}}^{a} L_\mathbf{2}^{a}(t) - 
    \gamma_\mathbf{m, 1}^{a} I_\mathbf{m, 1}^{a}(t), \\
I_\mathbf{m, 2}^{a\,\prime}(t) = {} & 
    \gamma_\mathbf{m, 1}^{a} I_\mathbf{m, 1}^{a}(t) - 
    \gamma_\mathbf{m, 2}^{a} I_\mathbf{m, 2}^{a}(t), \\
I_\mathbf{m, 3}^{a\,\prime}(t) = {} & 
    \gamma_\mathbf{m, 2}^{a} I_\mathbf{m, 2}^{a}(t) - 
    \gamma_\mathbf{m, 3}^{a} I_\mathbf{m, 3}^{a}(t), \\
I_\mathbf{p}^{a\,\prime}(t) = {} & 
    (1 - p^{a}) \alpha_{\mathbf{l}, \mathbf{2}}^{a} L_\mathbf{2}^{a}(t) - 
    \alpha_\mathbf{p}^{a} I_\mathbf{p}^{a}(t), \\
I_\mathbf{i, 1}^{a\,\prime}(t) = {} & 
    \alpha_\mathbf{p}^{a} I_\mathbf{p}^{a}(t) - 
    \gamma_\mathbf{i, 1}^{a} I_\mathbf{i, 1}^{a}(t), \\
I_\mathbf{i, 2}^{a\,\prime}(t) = {} & 
    \gamma_\mathbf{i, 1}^{a} I_\mathbf{i, 1}^{a}(t) - 
    \gamma_\mathbf{i, 2}^{a} I_\mathbf{i, 2}^{a}(t), \\
I_\mathbf{i, 3}^{a}(t) = {} & 
    \gamma_\mathbf{i, 2}^{a} I_\mathbf{i, 2}^{a}(t) - 
    \gamma_\mathbf{i, 3}^{a} I_\mathbf{i, 3}^{a}(t), \\
I_{\mathbf{h}}^{a\,\prime}(t) = {} & 
    h^{a} (1 - \xi^{a})  \gamma_\mathbf{i, 3}^{a} 
        I_\mathbf{i, 3}^{a}(t) - 
    \gamma_{\mathbf{h}}^{a} I_{\mathbf{h}}^{a}(t), \\
I_{\mathbf{c}}^{a\,\prime}(t) = {} & 
    h^{a} \xi^{a} \gamma_\mathbf{i, 3}^{a} 
        I_\mathbf{i, 3}^{a}(t) - 
    \gamma_{\mathbf{c}}^{a} I_{\mathbf{c}}^{a}(t), \\
I_{\mathbf{cr}}^{a\,\prime}(t) = {} & 
    (1 - \mu^{a}) 
        \gamma_{\mathbf{c}}^{a} I_{\mathbf{c}}^{a}(t) - 
    \gamma_{\mathbf{cr}}^{a} I_{\mathbf{cr}}^{a}(t), \\
R^{a\,\prime}(t) = {} & 
    \gamma_\mathbf{m, 3}^{a} I_\mathbf{m, 3}^{a}(t) + 
    (1 - h^{a}) 
      \gamma_\mathbf{i, 3}^{a} I_\mathbf{i, 3}^{a}(t) + 
    \gamma_{\mathbf{h}}^{a} I_{\mathbf{h}}^{a}(t) + 
    \gamma_{\mathbf{cr}}^{a} I_{\mathbf{cr}}^{a}(t), \\
D^{a\,\prime}(t) = {} & 
    \mu^{a} \gamma_{\mathbf{c}}^{a} I_{\mathbf{c}}^{a}(t), 
\end{split}
\end{equation}

where the index $a \in \{1, \ldots, 8\}$ represents the corresponding age group. 

Since we want to take into account the different characteristics of the disease in various age groups, we stratified the Hungarian population into eight groups, using the same age structure that we used in the questionnaires. The model parameters are calibrated based on comprehensive literature review and they are aligned to private data provided by the National Public Health Center in Hungary (for a previously published parameter set, see \cite{RostViruses}). Since the previously mentioned model was parameterized for seven age groups, we slightly changed the parameter vectors as shown in Table \ref{table:params-aged}. Actually, we added another aspect to the disease transmission term, which considers age-dependent susceptibility $\sigma$ of individuals: we set this value to $1.0$ except for the first two age groups, for which we use $0.5$. This aligns with the observations that children are less likely to get infected at the contact with an infected individual. 

{\renewcommand{\arraystretch}{1.0}
\begin{table}[h]
\centering
\begin{tabular}{|l|l|l|l|l|l|l|l|l|l|l|}
\hline
\multicolumn{2}{|l|}{\bf Probability / Age group} & 
\multicolumn{1}{|l|}{\bf $\mathbf{0}$--$\mathbf{4}$} & 
\multicolumn{1}{|l|}{\bf $\mathbf{5}$--$\mathbf{14}$} & 
\multicolumn{1}{|l|}{\bf $\mathbf{15}$--$\mathbf{29}$} & 
\multicolumn{1}{|l|}{\bf $\mathbf{30}$--$\mathbf{44}$} & 
\multicolumn{1}{|l|}{\bf $\mathbf{45}$--$\mathbf{59}$} & 
\multicolumn{1}{|l|}{\bf $\mathbf{60}$--$\mathbf{69}$} & 
\multicolumn{1}{|l|}{\bf $\mathbf{70}$--$\mathbf{79}$} & 
\multicolumn{1}{|l|}{\bf $\mathbf{80}$--} \\
\hline
Asymptomatic course & 
    $p^a$ &
    $0.9$ & $0.8$ & $0.7$ & $0.6$ & $0.4$ & $0.3$ & $0.2$ & $0.1$ \\
\hline
Hospitalization or & 
    {} & {} & {} & {} & {} & {} & {} & {} & {} \\
\quad intensive care (from $I_3^a$) & 
    $h^a$ &
    $0.00045$ & $0.00045$ & $0.0041$ & 
    $0.0028$ & $0.1094$ & $0.2529$ & $0.4663$ & $0.4965$ \\
\hline
Intensive care & 
    {} & {} & {} & {} & {} & {} & {} & {} & {} \\
\quad (given hospitalization) & 
    $\xi^a$ &
    $0.333$ & $0.333$ & $0.312$ & $0.297$ & 
    $0.292$ & $0.293$ & $0.293$ & $0.293$ \\
\hline
Fatal outcome & 
    {} & {} & {} & {} & {} & {} & {} & {} & {} \\
\quad (from $C_r^a$) & 
    $\mu^a$ &
    $0.2$ & $0.2$ & $0.216$ & 
    $0.25$ & $0.582$ & $0.678$ & $0.687$ & $0.7$ \\
\hline
Susceptibility parameter& 
    $\sigma^a$ &
    $0.5$ & $0.5$ & $1.0$ & $1.0$ & $1.0$ & $1.0$ & $1.0$ & $1.0$ \\
\hline
\end{tabular}
\caption{
    Age-dependent epidemiological parameters of COVID-19 for eight age groups
}
\label{table:params-aged}
\end{table}}

Simulating an epidemic model requires determining the epidemiological parameters along with initial state at the start of the simulation. In our analysis, we assume that for the first wave we do not have population-level epidemic spread, however, for the second wave we consider that $1\%$ of the population was infected during the first wave, and an additional $1\%$ was recovered from an outbreak started already in the summer, i.e. we already have infected individuals distributed over all age groups. The latter approach enables us to pass proper initial values for the deterministic model.

\section*{Effective reproduction number}

The transmission part of an epidemic model depends on the contact patterns between people in the susceptible and infected compartments, and the probability of virus transmission during the contacts, which can differ for different infectious compartments. This implies that we have to give estimates for $\beta^{(k, a)}_{X}$, which corresponds to the transmission rate of an infectious individual from $X$ and age group $k$ at contact with a susceptible from age group $a$, where $X\in\{P, M_1, M_2, M_3, I_1, I_2, I_3\}$. For this purpose, we compute the Next Generation Matrix (NGM) and baseline transmission rate $\beta_0$ using techniques of \cite{DiekmannHesterbeek}. Since the actually defined model only slightly differs from the one in \cite{RostViruses}, we might omit the detailed elaboration and highlight only the components have to be changed for the calculations, whereas the scheme of the computation remains the same.

For now, the block-diagonal transitional matrix $\Sigma$ has to be modified at elements $(3,2), (4,2), (4,3)$ and $(7,3)$, therefore we have
\begin{equation*}
\Sigma_a = 
\begin{bmatrix}
- \alpha_{\mathbf{l}, \mathbf{1}}^{a} & 
    0 & 0 & 0 & 0 & 0 & 0 & 0 & 0 \\
\alpha_{\mathbf{l}, \mathbf{1}}^{a} & -\alpha_{\mathbf{l}, \mathbf{2}}^{a} & 
    0 & 0 & 0 & 0 & 0 & 0 & 0 \\
0 & 
    (1 - p^{a}) \alpha_{\mathbf{l}, \mathbf{2}}^{a} & -\alpha_{\mathbf{p}}^{a} & 
    0 & 0 & 0 & 0 & 0 & 0 \\
0 & 
    p^{a} \alpha_{\mathbf{l}, \mathbf{2}}^{a} & 
    0 & 
        - \gamma_{\mathbf{m}, \mathbf{1}}^{a} & 
    0 & 0 & 0 & 0 & 0 \\
0 & 0 & 0 & 
    \gamma_{\mathbf{m}, \mathbf{1}}^{a} & -\gamma_{\mathbf{m}, \mathbf{2}}^{a} &
    0 & 0 & 0 & 0 \\
0 & 0 & 0 & 0 & 
    \gamma_{\mathbf{m}, \mathbf{2}}^{a} & -\gamma_{\mathbf{m}, \mathbf{3}}^{a} &
    0 & 0 & 0 \\
0 & 0 & 
    \alpha_{\mathbf{p}}^{a} & 
    0 & 0 & 0 & 
    - \gamma_{\mathbf{i}, \mathbf{1}}^{a} & 0 & 0 \\
0 & 0 & 0 & 0 & 0 & 0 & 
    \gamma_{\mathbf{i}, \mathbf{1}}^{a} & -\gamma_{\mathbf{i}, \mathbf{2}}^{a} &
    0 \\
0 & 0 & 0 & 0 & 0 & 0 & 0 & 
    \gamma_{\mathbf{i}, \mathbf{2}}^{a} & -\gamma_{\mathbf{i}, \mathbf{3}}^{a} \\
\end{bmatrix}
\label{eq:trnmtx}
\end{equation*}
for $a = 1, \ldots, 8$. 

The transmission matrix $\textbf{T}$ is affected by the above mentioned age-dependent susceptibility parameter for the first two age groups (since these parameters differ from 1). In these cases, we have to multiply all elements of the building blocks $\textbf{T}_1$ and $\textbf{T}_2$ with the respective susceptibility parameter value (that we assumed $0.5$ for both age groups, see discussion above).

Using NGM methodology, for a given initial contact matrix and observed $R_t$, we are able to estimate the baseline transmission rate finalizing the parametrization of the epidemic model. For calculating the effective reproduction number, on the one hand, we update the online measured contact matrix on a daily basis. On the other hand, for the second wave (when the virus spread across the whole country), we have to scale the elements of the contact matrices by the proportion of the susceptibles actually given by the model. We perform this as multiplying each column of the matrix by the respective proportion value. 

\subsection*{Seasonality effects}

\begin{figure}[!ht]
    \centering
    \includegraphics[width=0.6\textwidth]{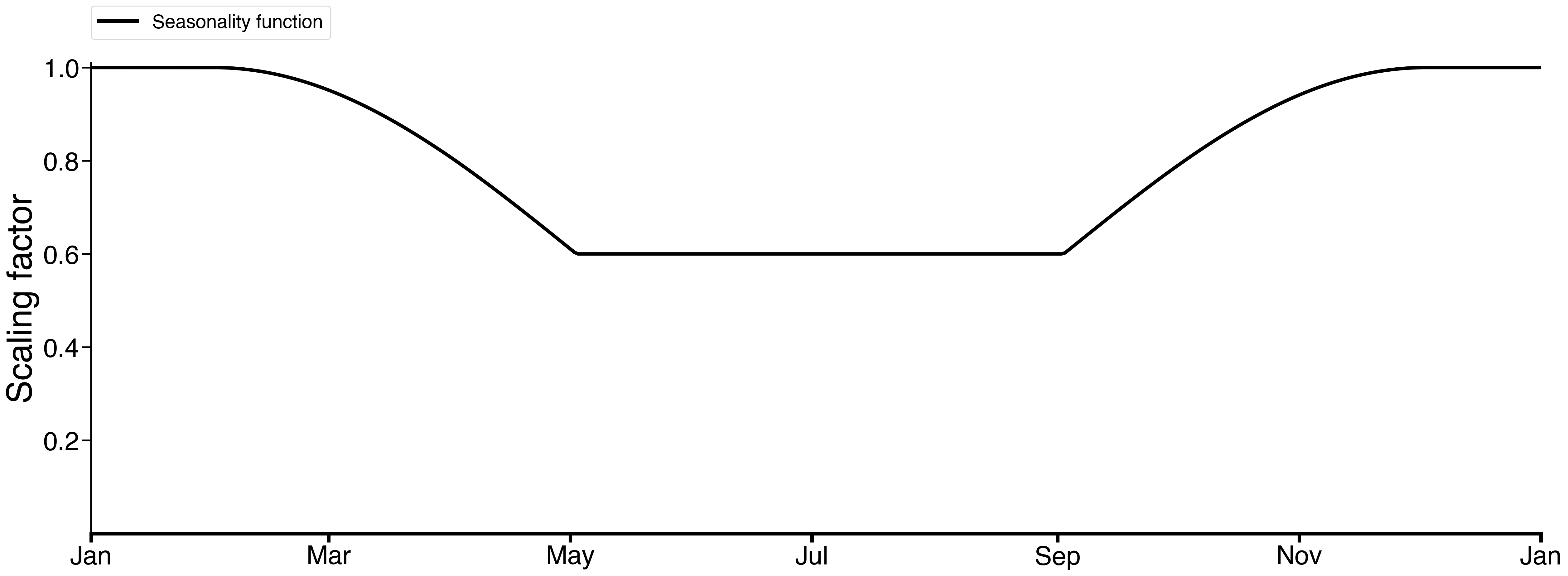}
    \caption{Seasonality function}
    \label{fig:seasonality}
\end{figure}

Since we investigate a nine-month period of 2020, we cannot neglect the effect of seasonal changes, which is incorporated into the model via scaling the baseline transmission rate by a time-varying function called seasonality function. This function is usually chosen for a 1-year periodic sine or cosine function \cite{RostViruses,balcan2009seasonal}, but our experimental observations from modeling the pandemic in Hungary show that the function shown in Figure~\ref{fig:seasonality} aligns better with the epidemic data. Clearly, during warmer periods of the year (from end of the spring until early autumn) the transmission rate is reduced in the population e.g. due to weather conditions and better natural ventilation, while a ramp-up and ramp-down phase are considered after and before this period, respectively. Furthermore during the winter and summer periods the efficiency rate does not change significantly, thus we assume that this rate is constant over these time intervals. Finally we set the ratio of 0.6 between summer and winter time~\cite{balcan2009seasonal,gozzi2022anatomy} and we kept the sinusoidal change in the complementary part for a year.

\end{document}